# Anti-Stokes excitation of solid-state quantum emitters for nanoscale thermometry

*Toan Trong Tran,[1,*] Blake Regan,[1] Evgeny A. Ekimov,[2] Zhao Mu,[3] Zhou Yu,[3] Weibo Gao,[3] Prineha Narang,[4] Alexander S. Solntsev,[1] Milos Toth,[1] Igor Aharonovich[1] and Carlo Bradac[1,*]*

[1]School of Mathematical and Physical Sciences, University of Technology Sydney, Ultimo, NSW, 2007, Australia
[2]Physics, RAS Kaluzhskoe Road 14, Troitsk, 142190, Russia
[3]Division of Physics and Applied Physics, School of Physical and Mathematical Sciences, Nanyang Technological University, 637371, Singapore
[4]John A. Paulson School of Engineering and Applied Sciences, Harvard University, Cambridge, MA, USA
[*]Corresponding authors: trongtoan.tran@uts.edu.au; carlo.bradac@uts.edu.au

**Abstract**
*Color centers in solids are the fundamental constituents of a plethora of applications such as lasers, light emitting diodes and sensors, as well as the foundation of advanced quantum information and communication technologies. Their photoluminescence properties are usually studied under Stokes excitation, in which the emitted photons are at a lower energy than the excitation ones. In this work, we explore the opposite Anti-Stokes process, where excitation is performed with lower energy photons. We report that the process is sufficiently efficient to excite even a single quantum system— namely the germanium-vacancy center in diamond. Consequently, we leverage the temperature-dependent, phonon-assisted mechanism to realize an all-optical nanoscale thermometry scheme that outperforms any homologous optical method employed to date. Our results frame a promising approach for exploring fundamental light-matter interactions in isolated quantum systems, and harness it towards the realization of practical nanoscale thermometry and sensing.*

Stokes and Anti-Stokes emission are fundamental phenomena widely used to study the physico-chemical and optical properties of materials. Stokes (Anti-Stokes) photoluminescence occurs when the energy of the emitted photons is lower (higher) than that of the absorbed ones (*1*). In the Anti-Stokes case, the extra energy that causes upconversion of the photons can be acquired through a variety of mechanisms, ranging from multi-photon absorption (*2*) to Auger recombination (*3*) and phonon absorption (*4*). The latter, relevant to this work, is illustrated in Figure 1A, B. A photon with energy $h\nu_{exc}$ at the long-wavelength tail of the absorption spectrum excites an electron from a thermally-populated first vibronic state ($n_0 = 1$) of the electronic ground state $E_0$, to the bottom manifold ($n_1 = 0$) of an excited electronic state $E_1$ [red arrow]. The system then returns to the ground state via spontaneous emission of an upconverted photon with a mean energy $h\nu_{se} > h\nu_{exc}$ [yellow arrow]. This phonon-assisted Anti-Stokes excitation process scales exponentially with temperature and is the bedrock of a variety of fundamental studies (e.g. cavity quantum electrodynamics (*5*)), as well as practical applications such as optical cryocooling (*6, 7*), bioimaging (*8*) and Raman spectroscopy (*9*). However, Anti-Stokes photoluminescence (PL) is inherently inefficient, and all work done to date on solid-state defects has been focused on ensembles (*10-12*) rather than individual point defects.

Here, we demonstrate that Anti-Stokes PL can be used to study isolated quantum systems—specifically atom-like color centers in diamond, over a large range of temperatures. We explore the mechanism for some of the most studied diamond defects, the nitrogen-vacancy (NV) (*13*) the silicon-vacancy (SiV) (*14, 15*) and the germanium-vacancy (GeV) (*16*) center. We show that Anti-Stokes excitation of selected diamond color center is an efficient process, detectable by standard photoluminescence spectroscopy and leverage this finding to demonstrate upconversion PL from a single, isolated GeV defect. We show that the Anti-Stokes excitation process is thermally-activated and proceeds through a phonon-photon absorption pathway rather than through multi-photon absorption. We exploit the high Anti-Stokes excitation efficiency to introduce an innovative approach for all-optical nanoscale thermometry based on the temperature-dependence of the Anti-Stokes to Stokes PL intensity ratio. Our technique outperforms all other previously reported all-optical nanothermometry methods.

To frame the scope of the Anti-Stokes process for quantum emitters and its capacity for developing nanoscale sensing applications, we characterized diamond samples (cf. Methods) containing germanium-vacancy (GeV), silicon-vacancy (SiV) and nitrogen-vacancy (NV) centers. A schematic illustration of a diamond defect in the split-vacancy configuration (i.e. GeV, SiV) is shown in Figure 1B. For each of the diamond defects, we selected a specific pair of excitation lasers (cf. Methods and SI, Fig. SI1) with energies above (Stokes) and below (Anti-Stokes) the zero-phonon line (ZPL) of each emitter. Figures 1C–E show room temperature Stokes [blue] and Anti-Stokes [ocher] PL spectra for ensembles of GeV, SiV and NV centers, respectively. Note that the sharp edges of the emission peaks are due to band-pass filters used to suppress the excitation lasers. The insets show the complete Stokes PL spectra of each color center. At room temperature, all color centers show Anti-Stokes PL. To confirm that the upconversion was not caused by multi-photon absorption, we measured photoluminescence intensity vs excitation power and concluded that the scaling does obey one-photon rather than two-photon absorption dynamics (cf. Supplementary Information, Fig. SI2).

Next, we established a direct, quantitative comparison amongst the Anti-Stokes to Stokes PL ratios of the studied centers. Normalizing the Anti-Stokes intensity makes the comparison independent of the density of defects amongst the different samples. The comparison does, nonetheless, issue some caveats. The first regards the selection of the Stokes and Anti-Stokes laser excitation wavelengths. Our hypothesis is that the Anti-Stokes excitation process involves vibronic states of the defects which are populated via the absorption of phonons by ground-state electrons. It therefore follows that the process is proportional to the phonon density of states, making Anti-Stokes absorption ideally the most efficient for excitation wavelengths matching the density maximum—and desirably not too narrow, spectrally. Simultaneously, for the comparison to be meaningful the difference between Anti-Stokes and Stokes excitation energies should be similar for the different color centers. Further, for practical sensing realizations, one must ultimately consider the number of color centers per unit volume of diamond realistically achievable for each type of defect—as this affects the signal-to-noise ratio and thus the resolution of the sensor.

Bearing these caveats in mind, we find that SiV and GeV centers outperform NV centers under our experimental conditions—their Anti-Stokes emission efficiency is higher as is their attainable density of defects per nanodiamond (*17, 18*). The measured Anti-Stokes to Stokes PL intensity ratios are similar for GeV and SiV centers, and approximately three orders of magnitude higher than that for NV centers: $I_{AS}/I_S|_{GeV} = (8.4\pm3.3)\times10^{-2}$, $I_{AS}/I_S|_{SiV} = (13.2\pm1.1)\times10^{-2}$ and $I_{AS}/I_S|_{NV} = (11.9\pm2.8)\times10^{-5}$. The lower efficiency of the Anti-Stokes process for the NV center is somewhat counterintuitive. The NV center displays a large phononic sideband, which trivially suggests more efficient coupling to the lattice and a more efficient Anti-Stokes excitation process compared to that of the spectrally-narrower SiV and GeV centers. The much lower value of the ratio $I_{AS}/I_S$ for the NV in our experiment is mainly due to the Anti-Stokes excitation laser being quite far below the NV ZPL energy (224.40 meV), at the long-wavelength tail of the phonon side band. Additional contributing factors to the low PL Anti-Stokes emission are the NV photo-ionization process (*19*) and the recently-proposed NV–N tunneling mechanism in nitrogen-rich diamond samples (*20*)—hinted by the difference between the Stokes and Anti-Stokes PL spectra seen in Figure 1E.

The SiV and GeV centers have similar Anti-Stokes emission efficiencies (normalized to their respective Stokes ones), making them both good candidates for Anti-Stokes quantum measurements and potential nanothermometry applications (cf. Supplementary Information, Fig. SI3). However, owing to the fact that its excited state decay is highly nonradiative (*21*), the SiV center possesses a lower luminescence quantum efficiency than the GeV (*16*). A high quantum efficiency is desirable for it maximizes the PL signal-to-noise ratio, which ultimately determines the temperature and spatial resolution in nanothermometry. We, therefore, selected the GeV center as our primary candidate for the remainder of this work.

We start by demonstrating that Anti-Stokes PL measurements are feasible down to a single quantum emitter (i.e. a single atom-like defect). Figure 2 shows the systematic analysis for the GeV center. Figure 2A is a 25×25 μm² confocal PL scan of a single crystal diamond where the bright spots are areas that have been implanted to induce the inclusion of GeV centers (cf. Methods). In the surveyed confocal scan, we isolated single GeV centers—like the one highlighted by a dashed red circle in Figure 2A. Figure 2B shows the PL measurement of the defect. The ZPL is clearly visible at 602 nm. The quantum nature of the emitter is shown by the second-order autocorrelation function $g^{(2)}(\tau)$ which has a zero-delay-time value $g^{(2)}(\tau = 0) < 0.5$ (not background corrected)—considered indicative of a single-photon emitter (Fig. 2C). Only the ZPL signal (shaded in blue in Fig. 2B) was used for the antibunching measurement.

Next, we carried out Anti-Stokes excitation of the identified GeV center. Remarkably, the process is efficient enough that Anti-Stokes emission from a single GeV defect can be detected in a standard PL measurement. Figure 2D shows the Anti-Stokes signal from the single GeV center from Figure 2A under laser excitation at a wavelength of 637 nm, 38 mW of power and a total acquisition time of 12 minutes. This result is notable on its own: it demonstrates, for the first time, Anti-Stokes PL from a single solid-state defect.

The high efficiency of the GeV Anti-Stokes PL process makes it a compelling candidate for all-optical nanothermometry (*22, 23*). To quantify the sensitivity, resolution and range of a potential nanothermometer, we characterized the Stokes and Anti-Stokes PL signals from a nanodiamond (~400 nm) hosting an ensemble of GeV centers (cf. Supplementary Information, Fig. SI4), as a function of temperature. The nanodiamond GeV ensemble had a room-temperature PL intensity of ~$10^6$ counts/s, measured under 532-nm (Stokes) laser excitation at 500 μW, after a 595–615-nm bandpass filter.

Figure 3A shows the results for the Anti-Stokes excitation analysis (also cf. Supplementary Information, Fig. SI5, SI6). The intensity of the Anti-Stokes emission exhibits Arrhenius-type exponential scaling with temperature. The data fits very well the equation $Ae^{-(E_a/k_BT)}$, with $k_B$ being the Boltzmann constant and $E_a$ the value for the activation energy fixed at 102.96 meV—which is the difference in energy between the Anti-Stokes excitation laser and the GeV ZPL. The Arrhenius-type dependence shows that the Anti-Stokes excitation process is thermally activated, supporting our hypothesis that the Anti-Stokes excitation of diamond color centers involves the absorption of phonons from the lattice.

Notably, the existence of an exponential dependence between Anti-Stokes PL intensity and temperature makes the mechanism ideal for high-sensitivity nanothermometry. For the purpose of realizing a practical sensor, we use the ratio between Anti-Stokes and Stokes PL as the experimental observable. The normalization makes the sensor independent of experimental specificities (e.g. loss of detected photons due to absorption or scattering in certain environments like living cells, or samples that change phase during a heating/cooling measurement). Figure 3B displays the Anti-Stokes to Stokes photoluminescence intensity ratio as a function of temperature, measured over the range 110–330 K. Over this range, the $I_{AS}/I_S$ ratio fits well the exponential function $a + be^{-[c/(T-T_0)]}$. The strong dependence on temperature is highly advantageous, as it translates to extremely high sensitivity—based on the standard definition of sensitivity, as an absolute quantity which describes the smallest amount of detectable change in a measurement. In fact, the thermometer sensitivity matches (or far exceed) that of any other all-optical method (Figure 3C), including techniques based on Raman spectroscopy which boast high sensitivity over a broad temperature range (*24*), but are not suitable for nanoscale thermometry because they suffer from limited spatial resolution.

In terms of temperature resolution, the performance of the nanothermometer we investigated is comparable with the current best all-optical-based methods (*25-28*) with a noise-floor temperature resolution of 455 mK·Hz$^{-½}$, at room temperature. Note that due to the exponential dependence of the $I_{AS}/I_S$ ratio with temperature the resolution worsens at low temperatures yet improves rapidly at high temperatures. Specifically, in the range 110–330 K, the measured temperature resolution is 2.494–0.420 K·Hz$^{-½}$. Unlike sensitivity, the resolution is a relative quantity and can be improved, for instance by selecting nanodiamond hosts with a higher density of color centers or by reducing the measurement bandwidth, i.e. increasing the integration time for the PL signal.

To complete the discussion, we benchmark the characteristics and performance of our nanothermometer against those of the current field's bests. The first factor is utility. Our approach is an all-optical, microwave-free nanothermonetry technique based on diamond color centers. Nanothermometers of this type (*25-28*) are broadly appealing because of their high spatial resolution, low noise floor (i.e. high temperature resolution), wide temperature range and broad applicability. The second metric is sensitivity, where all-optical nanothermometers often do not perform as well, for many rely on measuring the temperature-dependence on observables such as ZPL frequency (*25, 27, 28*) or amplitude (*29*) which vary weakly compared to the, demonstrated herein, Anti-Stokes to Stokes emission intensity ratio. We also note that techniques based on measuring PL intensity amplitude (rather than ratio), such as that of the NV center ZPL (*29*), have limited applicability because they suffer from a range of artifacts such as changes in photon scattering and absorption caused by changes in temperature of the measured sample.

Our approach is not compromised by any of these shortcomings. The Anti-Stokes to Stokes PL ratio in diamond color centers reaches temperature sensitivities that match those of Raman-based sensors, while retaining the desirable utility features of the methods based on photoluminescent nanodiamonds, including a ~few-nm spatial resolution, as it works on single color centers that are stable in sub-10 nm nanodiamonds (*17, 18*). Note also that the exponential scaling with temperature of the ratio $I_{AS}/I_S$ makes the resolution of our method increase rapidly at high temperature. This makes it desirable, for instance, for temperature sensing in high-power electronics (*30*)—in virtue as well of diamond color centers being able to withstand extremely high temperatures (>1000 K).

Figure 3C visually captures the superior performance of our approach against other nanothermometry schemes. The graph shows an absolute comparison by plotting the relative sensitivity of each technique as a function of temperature. We define the relative sensitivity as $(\partial O/\partial T)/O$ where $O$ is the measured observable (e.g. ZPL frequency, ZPL amplitude, etc.) The graph shows the relative sensitivity based on: *i)* our Anti-Stokes to Stokes PL intensity ratio, *ii)* the frequency shift of the GeV ZPL in our Stokes PL spectra, equivalent to *iii)* the same measurement reported in the literature (*27*), *iv)* the ZPL wavelength shift of the SnV (*28*) and *v)* SiV (*25*) diamond color centers and *vi)* the intensity change of the NV ZPL in diamond (*29*). The sensitivity of our technique is superior to that of any of these competitive methods; it matches (or slightly outperforms) the relative sensitivity benchmark of *vii)* the Anti-Stokes to Stokes emission intensity ratio of a sensor based on Raman spectroscopy (*24*). For reference, there is an entire family of nanothermometers (*31-33*) based on the temperature-varying properties of quantum dots (QDs)—Figure 3C shows an example based on *viii)* spectral shift (*31*)—yet these are often limited to a narrow temperature range (*32, 33*). The nanothermometry landscape also includes upconversion nanoparticles (UCNPs) (*34-37*). In some cases (*34*) they can reach sensitivities comparable to that of our approach, but they usually suffer from limited range of operative temperatures and/or by low quantum efficiency (i.e. low resolution).

In conclusion, we have demonstrated Anti-Stokes photoluminescence from a single atom-like defect in diamond, and leveraged the process to demonstrate a new variant of all-optical nanothermometry with unprecedented performance. Our approach forms a basis for fundamental studies of solid-state quantum systems via Anti-Stokes processes, and for novel non-invasive sensing technologies.

**Methods**
**Samples.** The NV-sample consisted in synthetic type Ib ND powders (MSY ≤0.1 μm; Microdiamant) purified by nitration in concentrated sulphuric and nitric acid ($H_2SO_4$-$HNO_3$), rinsed in deionized water, irradiated by a 3-MeV proton beam at a dose of ($1 \times 10^6$ ions per $cm^2$ and annealed in vacuum at 700 °C for 2 h to induce the formation of NV centers (Academia Sinica, Taipei Taiwan) (*38*) The measured NDs average size is (150.5 ± 23.3) nm.
The SiV-sample consisted in NDs synthesized using a microwave plasma chemical vapor deposition (MPCVD) system from detonation ND seeds (size 4-6 nm). The growth was carried out for 30 minutes in a gas mix of hydrogen:methane 100:1, at 900 W microwave power and 60 Torr pressure. The synthesized NDs had size ~ 0.3–1 μm.
For GeV centers we looked at different samples. The first consisted of GeV centers synthesized using a MPCVD method, whereby the germanium was introduced externally as a solid or vapor source. The sample for the single GeV color centers is a high-purity single crystal diamond from Element Six [N] <1 ppb implanted with germanium ions at 35 keV using a nanoFIB system (ionLINE, RAITH Nanofabrication) and an implantation dose of 100 $Ge^+$ ions per spot. The sample was subsequently annealed at 1000 °C for 30 minutes in high vacuum.
The second consisted in diamond nanoparticles hosting GeV color centers synthesized from mixtures of Adamantane, $C_{10}H_{14}$ (Sigma Aldrich, purity > 99%) with small amount of Tetraphenylgermanium $C_{24}H_{20}Ge$ (Sigma Aldrich, purity > 95.5 %) at 9 GPa and 1500-1700 K, as described elsewhere (*39, 40*). The concentration of Ge in the growth system was about 0.4% calculated relative to the carbon-germanium mixture, Ge/(Ge+C).
The third sample was a diamond membrane embedded with GeV color centers and was prepared as followings. The $GeO_2$ covered membrane was placed in a MPCVD chamber, along with a ~1 × 1 $mm^2$ piece of metallic germanium ~1 cm away. The conditions were: hydrogen/methane ratio of 100:1 at 60 Torr, microwave power of 900W for 10 minutes to fabricate a ~400-nm intrinsic diamond layer that contains GeV color centers. The diamond membranes were then flipped 180° and thinned by an inductive coupled plasma reactive ion etching (ICP-RIE) with argon, oxygen and SF6 etch (2:3:1), at a pressure of 45 mTorr, with a forward power of 500 W and 100 W, for the ICP and RIE respectively.
**Optical Characterization.** The samples were mounted on a three-dimensional piezo-stage (ANPx series, Attocube Inc.) in a lab-built, open-loop cryostat (adapted from a ST500 cryostat; Janis) with flowing liquid nitrogen. The temperature at the sample was controlled via a cryogenic temperature controller (335; Lakeshore). Optical access to the sample is through a thin quartz window; the lasers are focused via a high numerical-aperture objective (NA 0.9, 100×, TU Plan Fluor; Nikon), back-collected, spectrally filtered and sent to either a spectrometer (SR303I, mounted with a Newton DU920P CCD Camera; Andor) or a pair of avalanche photodiodes (SPCM-AQR-14; Perkin Elmer) in a Hanbury-Brown and Twiss interferometer configuration (*41*). Stokes/Anti-Stokes excitation was carried out with the following lasers.

GeV sample. Stokes excitation was carried out with a CW diode-pumped solid-state DPSS laser (SDL-532-200T, DreamLasers) at 532 nm. Anti-Stokes excitation was carried out with a TO-Can laser diode (HL63142DG, Thorlabs) at 637 nm. For the Anti-Stokes excitation on a single GeV defect (Figure 3D), a short-pass dichroic was used to increase the excitation and collection efficiency.

SiV sample. Stokes excitation was carried out with a TO-Can laser diode (HL63142DG, Thorlabs) at 637 nm. Anti-Stokes excitation was carried out with a CW Titanium:Sapphire (Ti:Sap) laser (SolsTis, M2 Inc.) at 770 nm.

NV sample. Stokes excitation was carried out with a CW diode-pumped solid-state DPSS laser (SDL-532-200T, DreamLasers) at 532 nm. Anti-Stokes excitation was carried out with a picosecond gain switched laser diode (PiL607X; PILAS) operating in CW at 675 nm and with a CW Titanium:Sapphire (Ti:Sap) laser (SolsTis, M2 Inc.) at 720 nm.

**Thermometry.** We measured the ZPL photoluminescence intensity under Stokes and Anti-Stokes excitation, and determined the relative Anti-Stokes PL efficiency for each defect, which ultimately limits the sensitivity of the nanothermometer.

**Acknowledgements**

Financial support from the Australian Research council (via DP180100077, DE180100810 and DE180100070), the Asian Office of Aerospace Research and Development grant FA2386-17-1-4064, the Office of Naval Research Global under grant number N62909-18-1-2025 are gratefully acknowledged. E. A. Ekimov is grateful to RFBR for the financial support under Grant No. 17-52-50075. The authors thank Marcus W. Doherty, Neil B. Manson and Jeff Reimers for useful discussions.

**Author contributions**

C.B., T.T.T., I.A. and M.T. conceived the project. B.R., E.A.E., Z.M., Z.Y., and W.G. fabricated the samples. T.T.T. designed and conducted the measurements with the assistance from C.B. Data analysis was conducted by T.T.T. and C.B. All authors discussed the results and co-wrote the manuscript.

# Figures

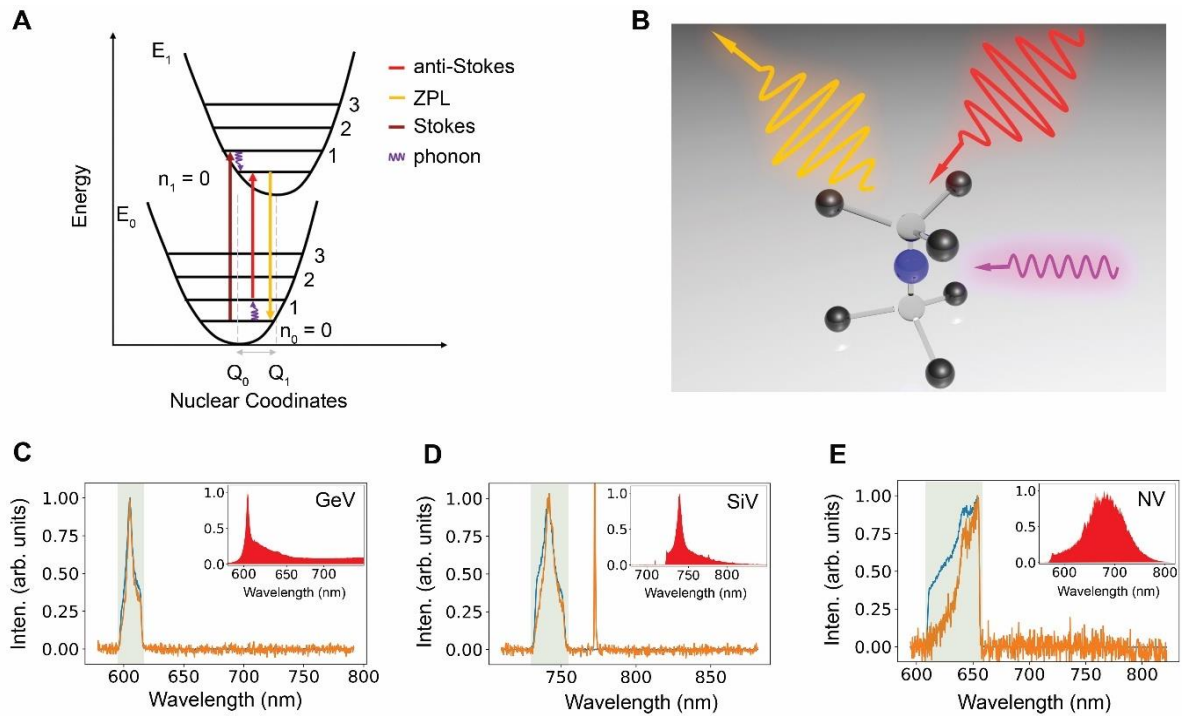

**Figure 1.** Stokes and Anti-Stokes luminescence processes for color centers in diamond. **A)** Energy diagram of representative electronic and vibrational energy levels for a diamond color center. The arrows show the lower (higher) energy of the Stokes (Anti-Stokes) photons with respect to the ZPL energy. In the Anti-Stokes case, the additional energy is acquired via phonon(s) absorption. **B)** artistic representation of the Anti-Stokes mechanism for a diamond color center which absorbs a lower-energy photon [wavy line, red] and emits a higher-energy one [wavy line, ocher] upon absorption of a phonon [wavy line, purple]. **C–E)** Photoluminescence spectra of the ZPL for nanodiamond GeV (C), SiV (D) and NV (E) centers under Stokes [blue] and Anti-Stokes [ocher] excitation (the full PL spectrum under Stokes excitation is shown in the relative inset). The ZPLs (605 nm for GeV, 739 nm for SiV and 639 nm for NV) are spectrally filtered by means of bandpass filters [semitransparent rectangular boxes]. For each measurement in (C), (D) and (E) the powers of the Stokes and Anti-Stokes excitation lasers are the same: PL intensities are normalized to unity for display purposes: the measured values for Anti-Stokes to Stokes PL intensity ratios for GeV, SiV and NV centers are $I_{AS}/I_S|_{GeV} = (8.4\pm3.3)\times10^{-2}$, $I_{AS}/I_S|_{SiV} = (13.2\pm1.1)\times10^{-2}$ and $I_{AS}/I_S|_{NV} = (11.9\pm2.8)\times10^{-5}$. (The line at ~770 nm in (D) is the Anti-Stokes excitation laser).

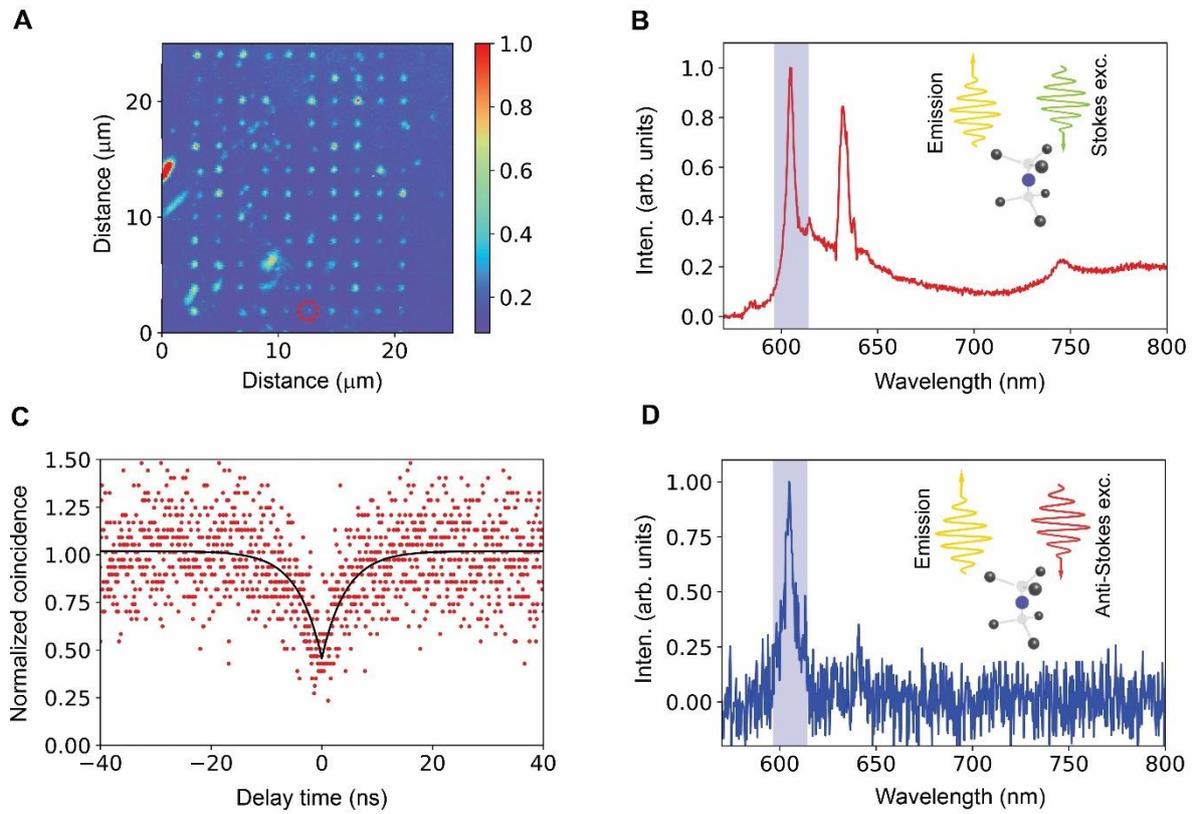

**Figure 2.** Characterization of Anti-Stokes emission from GeV color centers. **A)** Confocal image of a 25×25 μm² bulk diamond sample showing emission from GeV color centers. Each spot has a different density of GeVs. The spot indicated by the [red circle] is a single-photon GeV center as per analysis in (C). **B)** Photoluminescence spectrum acquired for the single center identified in (A). **C)** Second-order autocorrelation function $g^{(2)}(\tau)$ showing the sub-Poissonian statistic, at zero-delay time, indicative of a single photon source, $g^{(2)}(0) < 0.5$ (the value for $g^{(2)}(0)$ is not background-corrected). **D)** Anti-Stokes PL spectrum acquired from the single GeV center in (A). The acquisition was carried out for 12 minutes, with laser excitation at 637 nm and 38 mW of power. A bandpass filter (represented as a semi-transparent box around the ZPL of the spectrum) was used to acquire measurements in (C) and (D).

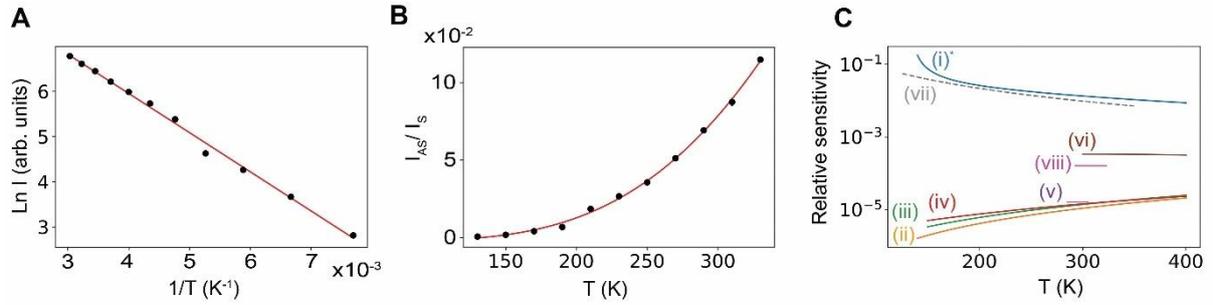

**Figure 3.** Characterization of the Anti-Stokes GeV-based nanothermometer. **A)** Temperature dependence of the PL intensity signal upon Anti-Stokes excitation (637-nm wavelength). The PL intensity was measured by monitoring the GeV's ZPL (605 nm) isolated with a bandpass filter. The data fit well the Arrhenius-type equation $Ae^{-(E_a/k_B T)}$, where the activation energy $E_a = 102.96$ meV is fixed to coincide with the difference in energy between the excitation laser and the germanium-vacancy's ZPL. **B)** Plot of the Anti-Stokes to Stokes PL ratio as a function of temperature. The ratio fits an exponential curve: $a + be^{-[c/(T-T_0)]}$, granting the method an extremely high sensitivity. The error bars of plots in (A) and (B) are represented as vertical, blue bars, and are mostly equivalent to or smaller than the size of the data points. **C)** Relative sensitivity plotted vs temperature for several different systems: our $I_{AS}/I_S|_{GeV}$ measurement *(i)\**, the frequency shift of the GeV ZPL in our Stokes PL spectra *(ii)*, and the equivalent measurement from the literature *(iii)*, the ZPL wavelength shift of the SnV *(iv)* and of the SiV center *(v)*, the intensity of the NV ZPL *(vi)*, the Raman $I_{AS}/I_S$ ratio achieved for a bulk thermometer *(vii)* and the spectral shift of quantum dots *(viii)*. The literature data are plotted over the entire temperature range demonstrated in each paper.